\documentclass[10pt]{iopart}
\usepackage{graphicx}
\usepackage{xcolor}
\usepackage[a4paper]{geometry}

\begin{document}

\title{Tunable Second-Order Structural Transition in As-Deficient MnAs}

\author{
	B D White$^{1,2,3}$,
	K Huang$^{2,3,4}$\footnote{
		Present Address: Lawrence Livermore National Laboratory, 7000 East Avenue, Livermore, CA 94550, USA
	},
	I L Fipps$^{2,3,4}$\footnote{
    Present Address: Quantum Design, 10307 Pacific Center Court, San Diego, CA  92121, USA
  },
	J J Hamlin$^{2,3}$\footnote{
    Present Address: Department of Physics, University of Florida, Gainesville, Florida 32611, USA
  },
	S Jang$^{2,3,4}$,
	G J Smith$^{5}$,
	B Xia$^{6}$,
	J W Simonson$^{5,6}$,
	C S Nelson$^{7}$,
	M C Aronson$^{5,8,9,10}$ and
  M B Maple$^{2,3,4}$
}
\address{$^{1}$Department of Physics, Central Washington University, 400 East University Way, Ellensburg, WA 98926-7442, USA}
\address{$^{2}$Department of Physics, University of California, San Diego, La Jolla, California 92093, USA}
\address{$^{3}$Center for Advanced Nanoscience, University of California, San Diego, La Jolla, California 92093, USA}
\address{$^{4}$Materials Science and Engineering Program, University of California, San Diego, La Jolla, California 92093, USA}
\address{$^{5}$Department of Physics and Astronomy, Stony Brook University, Stony Brook, NY 11794-3800, USA}
\address{$^{6}$Department of Physics, Farmingdale State College, Farmingdale, NY 11735-1021, USA}
\address{$^{7}$National Synchrotron Light Source II, Brookhaven National Laboratory, Upton, NY 11973}
\address{$^{8}$Condensed Matter Physics and Materials Science Department, Brookhaven National Laboratory, Upton, NY 11973-5000, USA}
\address{$^{9}$Stewart Blusson Quantum Matter Institute, University of British Columbia, 2355 East Mall, Vancouver, BC, V6T 1Z4, Canada}
\address{$^{10}$Department of Physics and Astronomy, University of British Columbia, 6224 Agricultural Road, Vancouver, B.C. V6T 1Z1, Canada}
\ead{mbmaple@ucsd.edu}

\begin{abstract}
	We report measurements of magnetization, specific heat, and thermal expansion performed on As-deficient MnAs single crystals (MnAs$_{0.968}$).
	Ferromagnetic order is observed near $T_C \simeq$ 306 K on warming and $T_C \simeq$ 302 K on cooling, which is consistent with previously-reported values for {stoichiometric MnAs samples}.
	In contrast, the second-order structural phase transition is observed at $T_S \simeq$ 353 K, which is nearly 50 K lower than in the stoichiometric compound.
	We observe differences in the thermal expansion of our samples when compared to reports of {stoichiometric} MnAs {including: (1) the $\sim$1.5\% volume decrease at $T_C$ is smaller than the expected value of 1.9\%, (2) the lattice parameters perpendicular to the basal plane exhibit a discontinuous jump of $\sim$1.1\% at $T_C$ instead of being continuous across $T_C$, and (3) thermal expansion perpendicular to the basal plane for $T_C \le T \le$ 315 K is negative rather than positive.}
	We also observe a correlation between the ratio of hexagonal lattice parameters, $c/a$, and $T_S$, strongly suggesting that the degree of structural anisotropy in MnAs could play an important role in tuning $T_S$.
\end{abstract}

\ioptwocol

\section{Introduction}
Ferromagnetic order is observed in the compound MnAs below a first-order phase transition at a Curie temperature of $T_C \simeq$ 317 K~\cite{Heusler04, Bean62}.
A concomitant structural phase transition from a hexagonal NiAs-type crystal structure with space group $P6_3/mmc$ below $T_C$ to an orthorhombic MnP-type structure with space group $Pnma$ above $T_C$ accompanies the magnetic phase transition~\cite{Wilson64}.
This magnetostructural phase transition in MnAs was the first experimentally-observed example of a first-order magnetic phase transition and the Bean and Rodbell model was developed to understand it~\cite{Bean62}.
An enormous magnetocaloric effect (MCE), characterized by large adiabatic temperature and isothermal entropic variations in response to applying a magnetic field, was observed near $T_C$ in MnAs~\cite{Campos06}.
This already appreciable MCE increases dramatically under applied pressure (with maximum effect at 2.23 kbar) to such an extent that the term colossal MCE was proposed to describe its magnitude~\cite{Gama04}.
Similar magnetocaloric properties were later observed at ambient pressure by applying chemical pressure in the system Mn$_{1-x}$Fe$_x$As ($0.003 \le x \le 0.0175$)~\cite{Campos06}.
These properties could make MnAs useful in applications involving magnetic refrigeration; though, the large thermal hysteresis exhibited by MnAs at $T_C$ limits its use in refrigeration systems~\cite{Campos06}.
Thin films of MnAs on GaAs substrate have also been studied for their potential applications in spintronics, magnetic tunneling junctions, and magneto-logic devices~\cite{Ryu08}.
In addition, the anomalous Hall conductivity has been measured in MnAs films on GaAs, raising the possibility of topological transport through band engineering~\cite{Helman2021}.
Computational results suggest that monolayers of MnAs would be half-metallic with a wide spin-gap of 3 eV and a Curie temperature of over 700 K, again suggesting possible applications in spintronics~\cite{Wang2019}.
Finally, intense interest in FeAs-based superconductors has led to searches for unconventional superconductivity in other arsenide compounds and MnAs and other binary arsenides have sometimes been studied as ``proxy'' structures for arsenide-based superconducting systems~\cite{Saparov12}.
As a result of this search, pressure-induced superconductivity has been observed in MnP and CrAs, which are isostructural with MnAs~\cite{Cheng15,Wu14}.

In addition to the well-studied magnetostructural phase transition at $T_C$, MnAs exhibits another structural phase transition near $T_S \simeq$ 400 K.
This second-order phase transition, which has received relatively little attention, involves a subtle distortion of the MnAs$_6$ octahedra.
It was recently suggested that this phase transition is driven by the relaxation of a soft phonon mode~\cite{Lazewski10, Lazewski11}, such that the hexagonal NiAs-type structure exhibited for $T < T_C$ is recovered for $T \ge T_S$.
Phonon dispersion results, calculated using spin-polarized density functional theory, suggest that a soft phonon mode is favorable for small unit-cell volumes~\cite{Lazewski10, Lazewski11}.
The phonon mode becomes soft as magnetoelastic coupling induces a significant decrease in the unit-cell volume upon warming through $T_C$, and the resulting ionic displacements lower the crystal symmetry from hexagonal to orthorhombic.
As temperature increases further, thermal expansion causes the volume of the unit cell to increase monotonically until $T_S$, where the energy associated with the soft phonon mode becomes unfavorably large, the ionic displacements relax, and a hexagonal crystal structure is recovered~\cite{Lazewski10, Lazewski11}.

In an effort to probe the conditions that set the temperature scale as well as the underlying mechanism that drives the phase transition at $T_S$, we studied the physical properties of As-deficient MnAs single crystals that exhibit an anomalously-low $T_S$ value.
Previous efforts to grow single crystals of MnAs have mainly involved Bridgman techniques~\cite{Barner77,Haneda77,Campos11}, extracting crystalline grains from stoichiometric elements melted together~\cite{Wilson64,Blois63JAP,Blois63PR}, or from zone refining~\cite{Paitz71}.
A recent study also reported successful synthesis of single crystals using a cubic-anvil high pressure/temperature method~\cite{Zhigadlo17}.
We were able to successfully synthesize single crystals via a molten flux technique.
The single crystals obtained by this method are relatively clean and undamaged, largely as a consequence of not needing to be cut from a large boule as in many of the other techniques used to synthesize single crystals of MnAs.
These single crystals were characterized by performing measurements of their physical properties including thermal expansion (measured with x-ray diffraction (XRD)), magnetization, and specific heat.
The second-order structural transition is observed near $T_S \simeq$ 353 K, which is nearly 50 K lower than is typically observed in polycrystalline samples.
The thermal expansion of these As-deficient MnAs single crystals, determined from the measured lattice parameters, quantitatively disagrees with reports of the thermal expansion of stoichiometric MnAs, which may be responsible for this anomalously-low value of $T_S$.
On the other hand, a strong correlation is demonstrated between the ratio of room-temperature hexagonal lattice parameter values, $c/a$, and $T_S$ as observed in our single crystals and other reports in the literature.
Based on these results, we propose that the degree of structural anisotropy could play an important role in tuning the transition temperature, $T_S$, of the second-order structural phase transition in MnAs.

\section{Experiment}
It is challenging to synthesize high-quality, stoichiometric single crystals of MnAs as a result of a combination of factors; these include the toxicity and high vapor pressure of sublimating As and a 1.9\% volume change at $T_C$ ~\cite{Willis54} that leads to the formation of severe macroscopic cracks and fractures upon cooling through $T_C$.
We were able to synthesize single crystals in a molten Sn flux.
The volatility of As was mitigated by using the binary precursor, InAs, so that As is ideally incorporated into the melt before it sublimates.
Mn powder (3N) and InAs (99.9999\%) were weighed in a 1:5 molar ratio and placed in an Al$_2$O$_3$ crucible with 99.999\% Sn pieces.
The molar ratio of Mn to Sn was 1:89.
Ta foil (99.95\%) 0.025 mm thick was used to cover the open end of the crucible.
Narrow slits were cut into the foil so that it could act as a sieve during centrifuging.
The crucible was double sealed in evacuated quartz ampoules.
The ampoules were heated from room temperature to 1000 $^{\circ}$C over 10 hrs and allowed to dwell at that temperature for 48 hrs.
Following slow cooling to 600 $^{\circ}$C over a period of 48 hrs (-8.3 $^{\circ}$C/hr), the ampoule was centrifuged to remove both the molten Sn flux and molten In.
The as-grown single crystals were etched in a dilute solution of HCl to remove any residual Sn flux on their surfaces.

\begin{figure}
\begin{center}
\includegraphics[width=\columnwidth]{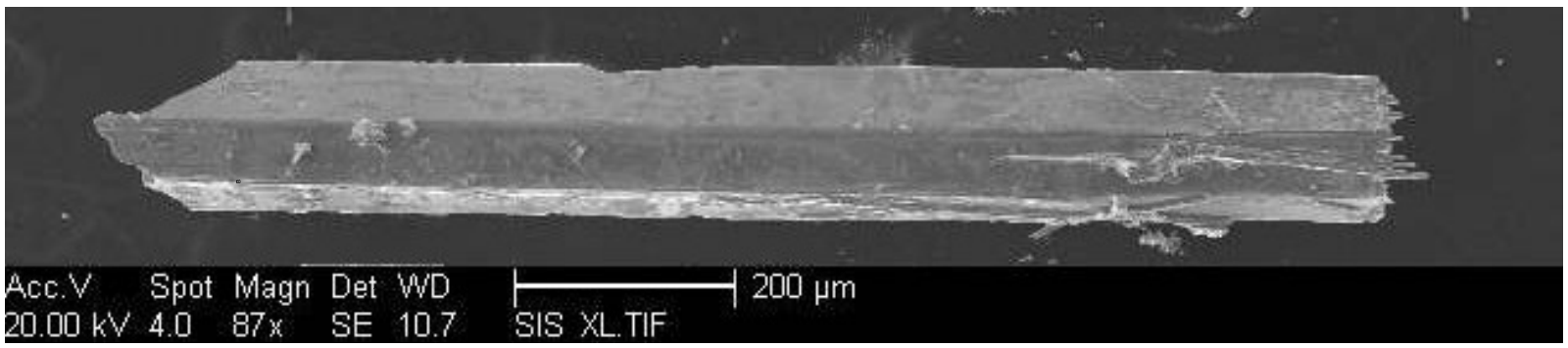}
\caption{\label{SEM_Image} Scanning electron microscope image of single crystal with 200 $\mu$m scale shown explicitly.
The needle-shaped crystal has a hexagonal cross-sectional morphology with the long direction parallel to the crystallographic $c$ axis.}
\end{center}
\end{figure}

The single crystals assume a needle-shaped morphology with the crystallographic $c$ axis aligned parallel to the long axis of the needle.
Typical lengths of 2-4 mm were obtained with the short dimensions (within the basal plane) being roughly 125 $\mu$m $\times$ 125 $\mu$m.
This morphology is similar to that reported for micro-crystals picked from the product of slow cooling stoichiometric melts~\cite{Blois63PR}.
The scanning electron microscope image of a representative single crystal in Fig.~\ref{SEM_Image} demonstrates that the needle has a hexagonal cross-sectional morphology.
Three distinct faces are visible in the image with three more hidden beneath the crystal.

Single crystal x-ray diffraction (XRD) measurements were carried out at 297 K using an Oxford Gemini single crystal diffractometer with an ATLAS detector and Cu-$K_{\alpha}$ radiation.
The refined solution was obtained via a charge-flipping algorithm~\cite{Petricek14, Palatinus07} from analysis of 1419 reflections after analytical absorption correction from a multi-faceted crystal.

The temperature dependencies of the lattice parameters were determined by four-circle high-resolution diffraction experiments carried out at beamline X21 of the National Synchrotron Light Source (NSLS) using an x-ray energy of 10 keV ($\lambda$ = 1.23985
\AA).
The apparatus was aligned for triple-axis crystal geometry, and the primary and analyzing crystals were respectively Si(111) and LiF (200) aligned for the diffraction condition.
The beam size was 1 $\times$ 1 mm$^2$.

The crystals were oriented by identifying the hard ([001] axis) and easy axes (within the basal plane) of single crystals via magnetization measurements.
Measurements of specific heat were performed in a physical properties measurement system (PPMS) DynaCool from Quantum Design (QD).
These measurements used a thermal-relaxation technique and were performed using Apiezon N-grease for temperatures $T < 300$ K and Apiezon H-grease for $T >$ 300 K.
Magnetization measurements were performed as a function of both temperature $T$ and magnetic field $H$ using a QD magnetic properties measurement system (MPMS).
Due to the small size of the single crystals, measurements of heat capacity and magnetization were performed on mosaics of aligned crystals.

\section{Results and discussion}

\subsection{Sample characterization}


The quality and phase-purity of the single crystals were assessed by performing refinements on single crystal XRD measurements.
The expected crystal structure at room temperature (NiAs-type structure with hexagonal space group $P6_3/mmc$) was observed and the lattice parameters were found to be similar to previous results for polycrystals~\cite{Saparov12} and single crystals~\cite{Campos11}.
A summary of the crystallographic parameters obtained from our refinement ($R_{obs}$ = 4.07) are found in Table~\ref{Table_Refinement}.
\begin{table}
\begin{center}
\caption{\label{Table_Refinement} Summary of crystallographic parameters resulting from refinement of single crystal x-ray diffraction measurements.}
\begin{tabular}{cc}
\hline
\hline
$a$ (\AA)           & 3.7172(3)                        \\
$b$ (\AA)           & 3.7172(3)                        \\
$c$ (\AA)           & 5.7044(4)                        \\
$V$ (\AA$^3$)       & 68.261(9)                        \\
\hline
Mn                  & (0,0,0), (0,0,1/2)               \\
occupancy           & 1                                \\
$U_{ani}$ (\AA$^2$) & 0.0107(15)                       \\
\hline
As                  & (1/3, 2/3, 1/4), (2/3, 1/3, 3/4) \\
occupancy           & 0.968(15)                        \\
$U_{ani}$ (\AA$^2$) & 0.081(13)                        \\
\hline
\hline
$R_{obs}$ = 4.07
\end{tabular}
\end{center}
\end{table}

\begin{figure}
\begin{center}
\includegraphics[width=\columnwidth]{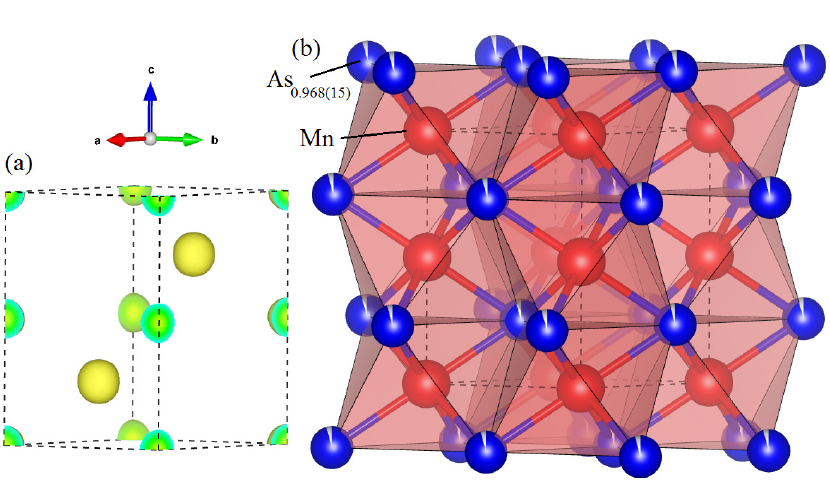}
\caption{\label{charge_density} (Color online) (a) The observed charge density that is generated by a charge flipping algorithm from single crystal x-ray diffraction (XRD) data is displayed within the unit cell of MnAs$_{0.968(15)}$.
A significantly larger charge density is observed on the As sites in the interior of the unit cell than on the Mn sites located on the boundaries of the unit cell.
(b) The crystal structure of our MnAs$_{0.968(15)}$ single crystals (as obtained from refinements of the single crystal XRD data) is displayed.
Blue spheres represent As while red spheres indicate Mn.
Dashed lines indicate the unit cell boundaries.}
\end{center}
\end{figure}

Energy dispersive x-ray spectroscopy (EDX) measurements were performed at multiple locations on each of several single crystals confirming that In and Sn impurities could be excluded to the 1\% level.
To perform additional chemical characterization, we studied the raw charge density using a charge flipping algorithm and our single crystal XRD data.
The results are displayed in Fig.~\ref{charge_density}(a).
Two different concentrations of charge density are evident, a relatively smaller one at the (0, 0, 0) and (0, 0, 1/2) positions and a relatively larger one at the (1/3, 2/3, 1/4) and (2/3, 1/3, 3/4) positions.
Figure~\ref{charge_density}(b) displays the crystal structure of our single crystals as determined from the refinement, revealing that the larger charge density is found on As sites and the lower charge density is located on Mn sites.
If the smaller charge density shown in Fig.~\ref{charge_density}(a) is fixed to represent full Mn occupancy, the larger density corresponds to only 96.8(15)\% of the expected As charge, indicating that this site is likely less than fully occupied. Since the In and Sn constituents of the flux are heavier than As, finding less than full occupancy of the As site suggests that co-occupancy by either impurity species is unlikely.
An alternative interpretation would be that the As site consist of (As + Sn/In + more than 3.2\% unoccupied).
Because x-ray scattering cross-sections are proportional to $Z^2$, neutron diffraction measurements would be necessary to rule this scenario out with absolute confidence.
Permitting fractional In or Sn occupancy of the As site did not improve the refinement of our XRD data, so we accept MnAs$_{0.968(15)}$ as the simplest model that explains the observed intensities.
Loss of As during the growth process is certainly plausible given its well-known volatility.

\subsection{Basic characterization of physical properties}

\begin{figure*}
\begin{center}
\includegraphics[width=\textwidth]{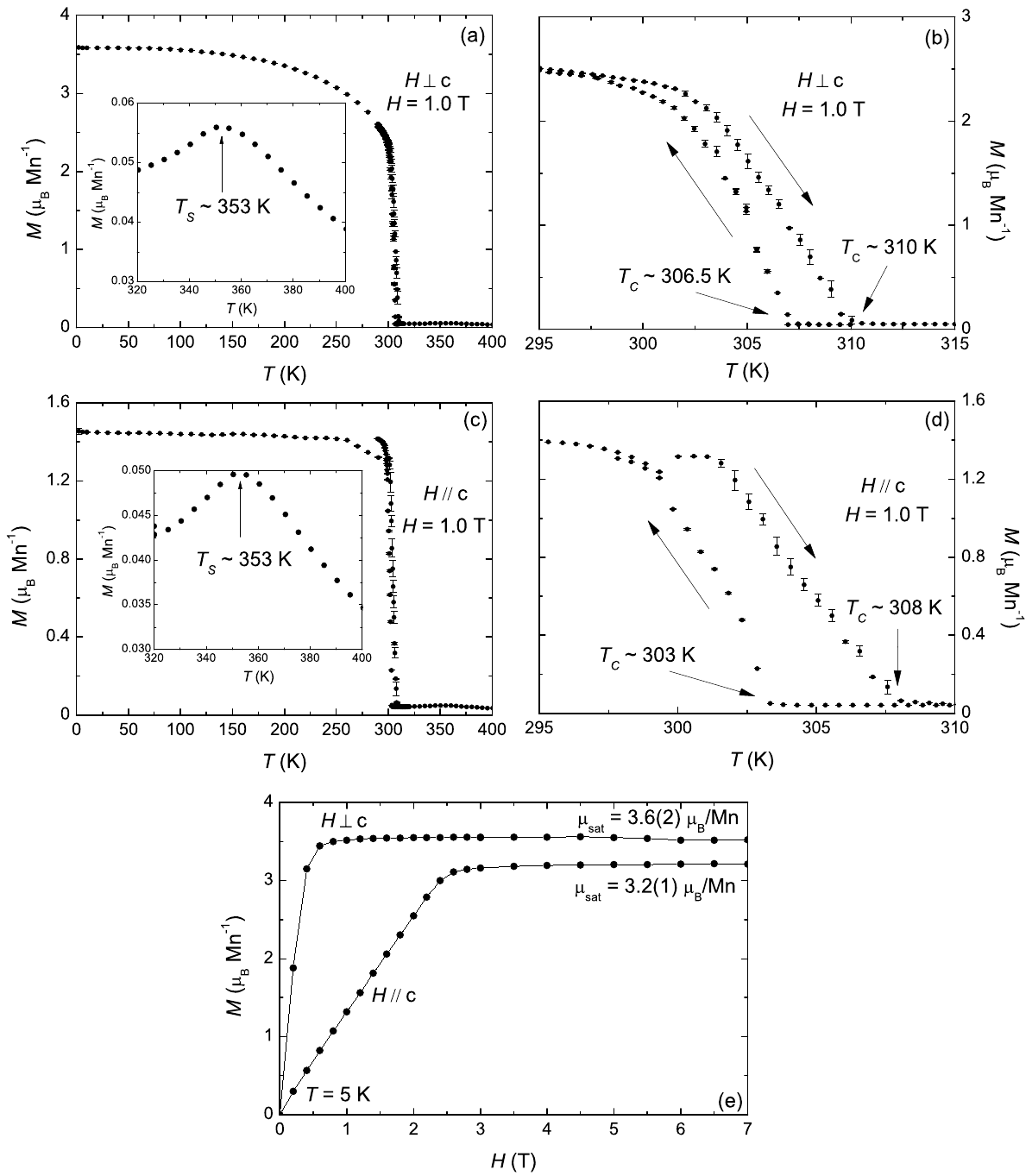}
\caption{\label{Magnetization} Magnetization, $M(T)$, measured in magnetic field $H$ = 1 T applied perpendicular ($H\perp c$) to the hexagonal structure's crystallographic $c$ axis (a)-(b) and parallel ($H\parallel c$) to the $c$ axis (c)-(d).
Hysteresis is observed at $T_C$ with a width of $\Delta T_C \simeq$ 4 K - 5 K in (b) and (d) where arrows explicitly designate data measured on warming and cooling.
The insets of (a) and (c) highlight the structural phase transition at $T_S \simeq 353$ K. (e) $M(H)$ measured at 5 K for $H\perp c$ and $H\parallel c$.
Lines are guides to the eye.}
\end{center}
\end{figure*}

Magnetization, $M(T)$, was measured in an applied magnetic field $H$ = 1 T and is displayed in Figs.~\ref{Magnetization}(a)-(b) for $H\perp c$ and Figs.~\ref{Magnetization}(c)-(d) for $H\parallel c$ (where $c$ is defined within the hexagonal structure).
A sharp ferromagnetic transition is observed just above 300 K in the main panels of Figs.~\ref{Magnetization}(a) and~\ref{Magnetization}(c).
The transition region is highlighted for $H\perp c$ in Fig.~\ref{Magnetization}(b) and for $H\parallel c$ in Fig.~\ref{Magnetization}(d).
The observed hysteresis is consistent with the transition's well-established first-order character~\cite{Bean62}.
If we define the temperature where the spontaneous magnetic order is first observed as $T_C$, then for $H\perp c$, $T_C \sim$ 310 K and $T_C \sim$ 306.5 K on warming and cooling through the transition, respectively.
For $H\parallel c$, the transition temperatures are $T_C \sim$ 308 K and $T_C \sim$ 303 K on warming and cooling, respectively.
The dependence of $T_C$ on magnetic field is known to be anisotropic~\cite{Blois63JAP}, so it is not surprising that the $T_C$ values we obtained at $H$ = 1 T are different for $H$ applied along the two distinct orientations.
The dependence of $T_C$ on $H$ for both $H\parallel c$ and $H\perp c$ is published in Ref.~~\cite{Blois63JAP}.
Using the experimental results for $dT_C/dH$ therein, we estimated what our $T_C$ values should be at $H = 0$ T, obtaining consistent values for both orientations of $T_C \simeq$ 306 K for warming and $T_C \simeq$ 302 K for cooling.
The size of the phase transition's thermal hysteresis (approximately $\Delta T_C \simeq$ 4 K - 5 K in each orientation) is smaller than the thermal hysteresis widths of $\Delta T_C \sim$ 7 K - 10 K that are typically reported for stoichiometric MnAs (see Table~\ref{Table}).
It has been previously suggested that a value of $\Delta T_C =$ 5 K might indicate {sample imperfections}~\cite{Blois63PR}, but a similar thermal hysteresis width has been reported for other single crystalline samples~\cite{Blois63JAP}.
{In this case, the smaller hysteresis width could reasonably be associated with the As-deficiency in our single crystals.}

The behavior of $M(T)$ below $T_C$ is anisotropic.
For $H\parallel c$, $M(T)$ jumps sharply to $\sim1.40~\mu_B$ per Mn ion at $T_C$, but is nearly temperature-independent below $T_C$ as seen in Fig.~\ref{Magnetization}(c).
In contrast, $M(T)$ jumps discontinuously to $\sim$2.35~$\mu_B$ per Mn ion at $T_C$ for $H\perp c$ and increases with decreasing temperature in a manner that is highly reminiscent of the behavior calculated using the model of Bean and Rodbell~\cite{Bean62}.
This model uses a molecular field theory to describe the ferromagnetic phase transition of materials with non-negligible magnetoelastic coupling and it predicts that $M(T)$ should exhibit a discontinuous increase at $T_C$ to $\sim$65\% of its zero-temperature value as MnAs enters its ferromagnetic state~\cite{Bean62}.
This appears to describe the magnitude of the jump in data measured for $H\perp c$ very well (see Fig.~\ref{Magnetization}(a)).

$M(T)$ data near the second-order structural phase transition at $T_S$ is highlighted in the insets of Figs.~\ref{Magnetization}(a) and~\ref{Magnetization}(c).
A feature is observed near $T_S \simeq$ 353 K for both orientations that exhibits the same peak-like character in each case (similar to published data~\cite{Campos11} measured in $H$ = 5 T).
Surprisingly, this value for $T_S$ is $\sim$50 K lower than that typically observed in stoichiometric MnAs (see Table~\ref{Table}).

Measurements of $M(H)$, performed at 5 K, are displayed in Fig.~\ref{Magnetization}(e) for both orientations.
Significant magnetocrystalline anisotropy has been previously reported in MnAs~\cite{Blois63JAP,Blois63PR}, with the hexagonal $c$ axis being identified as the hard direction~\cite{Haneda77,Blois63JAP,Blois63PR}.
This is confirmed in our measurements, which provide saturation moments of $\mu_{\mathrm{sat}}$ = 3.6(2)~$\mu_B$/Mn ion and $\mu_{\mathrm{sat}}$ = 3.2(1)~$\mu_B$/Mn ion for $H\perp c$ and $H\parallel c$, respectively.
The experimental uncertainties of $\mu_{\mathrm{sat}}$ are dominated by the resolution with which we could measure the mass of the mosaic of small single crystals used in the measurement of $M(H)$.
These values for $\mu_{\mathrm{sat}}$ agree reasonably well with previously reported values as seen in Table~\ref{Table}.
For $H\perp c$, $M(H)$ rapidly saturates near $H \sim$ 1 T, whereas $M(H)$ saturates near $H \sim$ 3 T for $H\parallel c$.
For $H < 3$ T and $H\parallel c$, $M(H)$ is quite linear.
Such behavior is reminiscent of $M(H)$ data for elemental Co~\cite{Honda26, Kaya28, Paige84}, which also exhibits a hexagonal crystal symmetry.
\begin{table*}
\begin{center}
\caption{\label{Table} Summary of properties for our MnAs$_{0.968(15)}$ single crystals and MnAs samples reported in other studies.
Note that the reported ratio of hexagonal lattice parameters, $c/a$, is at room temperature.}
\begin{tabular}{cccccccc}
\hline
\hline
Sample Details & $T_C$ (K) & $T_C$ (K) & $\Delta T_C$ (K) & $\mu_{\mathrm{sat}}$ ($\mu_B$/Mn) & $T_S$ (K) & $c/a$ & Reference\\
& warming & cooling & & & & & \\
\hline
Single crystal & 306       & 302   & 4   & 3.2(1), 3.6(2) at 5 K & \textbf{353-359}   & 1.5346 & This study \\
Single crystal & 311.65(5) & 307.1 & 4.6 &                       &           &        & ~\cite{Blois63JAP} \\
Polycrystal    & 312       & 305   & 7   &                       & $\sim$403 & 1.532  & ~\cite{Willis54} \\
Single crystal & 315.5     & 308   & 7.5 & 3.47 at 4 K           & 394       & 1.5330 & ~\cite{Campos11} \\
Polycrystal    & 316       &       &     & 3.6 at 2 K            & 384       & 1.5335 & ~\cite{Saparov12} \\
Polycrystal    & 317       & 306   & 11  & 3.3(1) at 4.2 K       & 390(5)    & 1.532  & ~\cite{Zieba78} \\
Polycrystal    & 318       & 311   & 7   & 3.2 at 4 K            &           &        & ~\cite{Gama04} \\
Single crystal & 318       & 305   & 13  & 3.40(3) at 0 K        & $\sim$400 & 1.533  & ~\cite{Haneda77}\\
\hline
\hline
\end{tabular}
\end{center}
\end{table*}

\begin{figure}
\begin{center}
\includegraphics[width=\columnwidth]{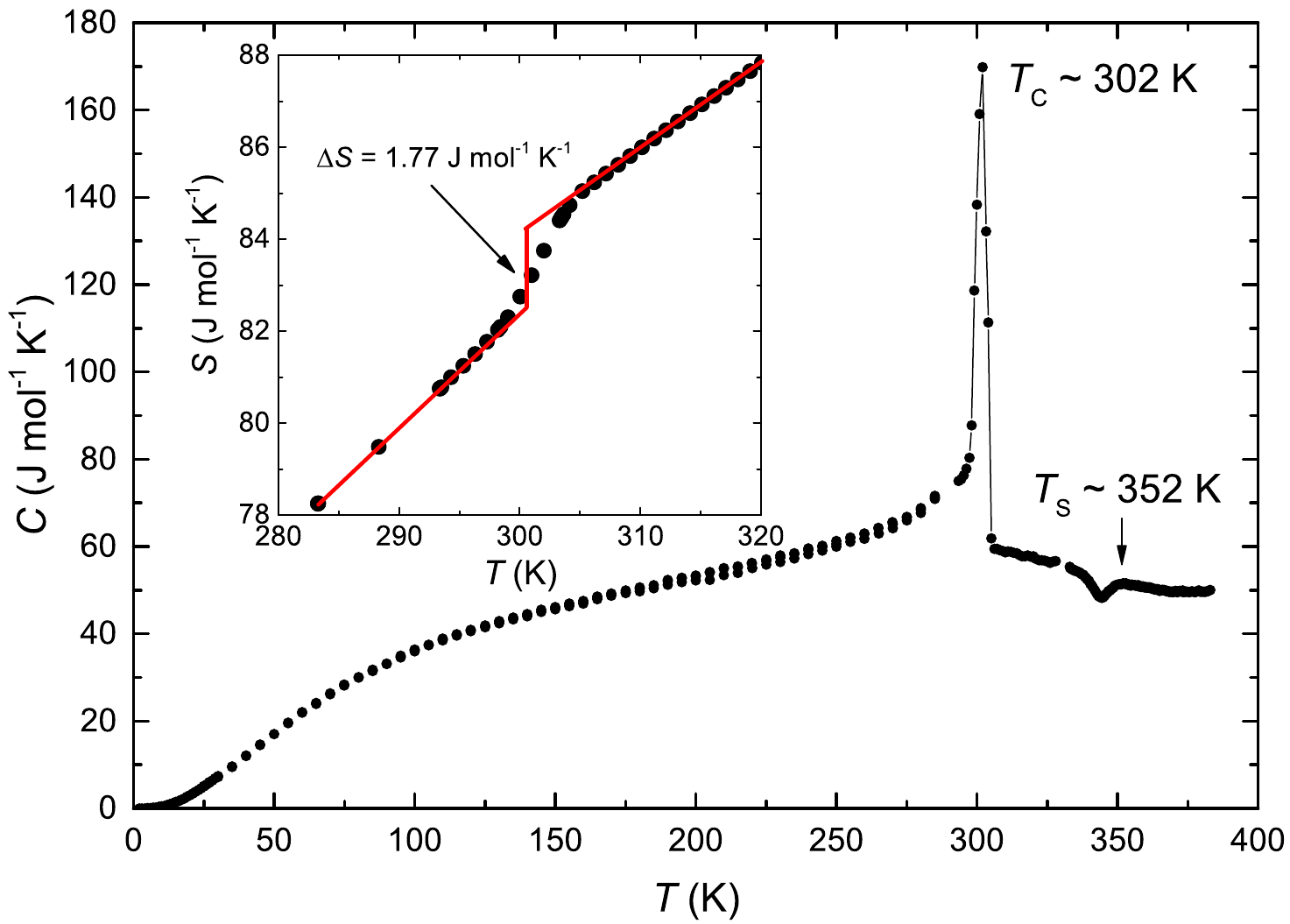}
\caption{\label{Heat_Capacity} The specific heat $C$ vs. temperature $T$ is displayed.
A sharp feature is observed at $T_C \simeq$ 302 K, which is associated with the first-order magnetostructural phase transition.
A smaller feature near $T_S \simeq$ 352 K is associated with the second-order structural phase transition.
The inset displays the specific entropy, $s$, as a function of temperature near $T_C$; it was determined by numerically integrating $C/T$ data with respect to $T$.
We estimated the discontinuity at $T_C$ to be $\Delta s$ = 1.77 J mol$^{-1}$ K$^{-1}$.}
\end{center}
\end{figure}

Specific heat $C(T)$ data are displayed in Fig.~\ref{Heat_Capacity}.
The first-order magnetostructural phase transition appears as a prominent feature near $T_C \simeq$ 302 K and a smaller feature associated with the second-order structural phase transition is observed near $T_S \simeq$ 352 K.
Both of these transition temperatures are consistent with the $T_C$ and $T_S$ values observed in measurements of $M(T)$.
The character of $C(T)$ is similar to an earlier report~\cite{Krokoszinski82} of measurements on a polycrystalline sample performed from low temperature to temperatures in excess of 400 K.

Table \ref{Table} displays the properties of MnAs as reported in this and other studies.
The values of $T_C$, $\Delta T_C$, and $\mu_{\mathrm{sat}}$ we obtain from our measurements are in relatively good agreement with other reported values.
In contrast, the value of $T_S \simeq$ 353 K we obtain for our single crystals is \textit{significantly} lower than other reported phase transition temperatures.

\begin{figure*}
\begin{center}
\includegraphics[width=\textwidth]{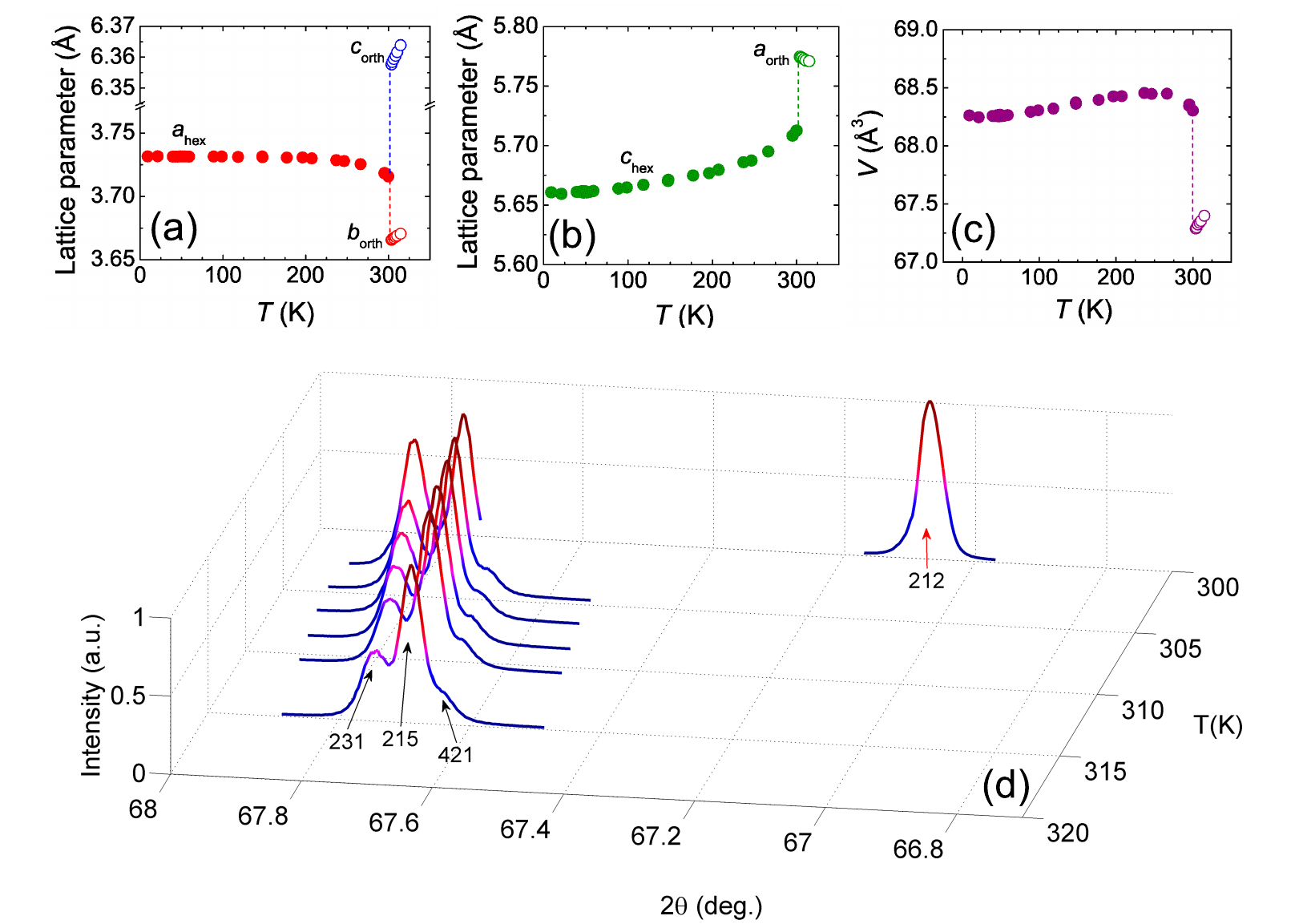}  
\caption{\label{Lattice_Parameters} (Color online) (a) The temperature dependence of the hexagonal lattice parameter $a_{\mathrm{hex}}$ for temperatures $T < T_C$ (filled circles) splits into the orthorhombic lattice parameters $b_{\mathrm{orth}}$ and $c_{\mathrm{orth}}$ for $T > T_C$ (unfilled circles).
A discontinuous jump is observed at $T_C$ as emphasized by the dashed lines. (b) The temperature dependence of the lattice parameter $c_{\mathrm{hex}}$ is plotted for $T < T_C$ (filled circles) and $a_{\mathrm{orth}}$ is shown for $T > T_C$ (unfilled circles).
A discontinuity is observed at $T_C$ as emphasized by the dashed line, indicating that the out-of-basal-plane thermal expansion is also discontinuous across $T_C$. (c) The hexagonal unit-cell volume, $V$, is plotted for $T < T_C$ (filled circles).
For $T > T_C$, half the orthorhombic unit-cell volume is plotted (unfilled circles) as described in the text.
$V$ exhibits a discontinuity at $T_C$, which is emphasized by a dashed line. (d) The hexagonal (212) Bragg reflection, which vanishes at $T_C$, is shown at a temperature (corresponds to the right scale) near $T_C$.
The orthorhombic (231), (215), and (421) Bragg reflections emerge for $T > T_C$ and their evolution with $T$ is shown as a function of $2\theta$.}
\end{center}
\end{figure*}

XRD measurements were performed on a representative single crystal at the National Synchrotron Light Source (NSLS) in order to measure the evolution of the lattice and crystal structure as a function of temperature.
In Figs.~\ref{Lattice_Parameters}(a) and \ref{Lattice_Parameters}(b), the hexagonal lattice parameters $a_{\mathrm{hex}}$ and $c_{\mathrm{hex}}$ are plotted as a function of temperature for $T < T_C$ and the orthorhombic lattice parameters $a_{\mathrm{orth}}$, $b_{\mathrm{orth}}$, and $c_{\mathrm{orth}}$ are plotted for $T > T_C$.
We note that these measurements were performed on warming.
Wilson and Kasper described~\cite{Wilson64} the relationship between the hexagonal and orthorhombic unit cells via the transformation: $a_{\mathrm{orth}}$ = $c_{\mathrm{hex}}$, $b_{\mathrm{orth}}$ = $a_{\mathrm{hex}}$, and $c_{\mathrm{orth}}$ = $\sqrt{3}a_{\mathrm{hex}}$.
The orthorhombic structure ($T_C \le T < T_S$) is essentially orthohexagonal with an axial ratio of $\sqrt{3}$.
In previous XRD measurements of polycrystalline samples, the thermal expansion has been shown to be quite anisotropic~\cite{Willis54}.
A large, discontinuous volume change in MnAs was observed at $T_C$, which occurs completely within the basal plane ($a_{\mathrm{hex}}$ is $\sim$0.9\% larger than $b_{\mathrm{orth}}$ at $T_C$).
On the other hand, the thermal expansion perpendicular to the basal plane is reported to be continuous across $T_C$~\cite{Willis54}.
The thermal expansion we measured for our single crystals within the basal plane and perpendicular to the basal plane, respectively, is shown in Figs.~\ref{Lattice_Parameters}(a) and \ref{Lattice_Parameters}(b).
The discontinuous jump in the basal plane at $T_C$, calculated to be $\sim$1.3\% by comparing $a_{\mathrm{hex}}$ and $b_{\mathrm{orth}}$ values at $T_C$ (see Fig.~\ref{Lattice_Parameters}(a)), is larger than the previously reported value from polycrystalline samples.
Another more significant difference is that there is a $\sim$1.1\% increase at $T_C$ between $c_{\mathrm{hex}}$ and $a_{\mathrm{orth}}$ (see Fig.~\ref{Lattice_Parameters}(b)), in contrast to reports~\cite{Willis54} of a continuous thermal expansion across $T_C$ perpendicular to the basal plane.
It is also notable that negative thermal expansion is observed in the orthorhombic phase of our single crystals ($T_C \le T < T_S$) perpendicular to the hexagonal basal plane (see $a_{\mathrm{orth}}(T)$ in Fig.~\ref{Lattice_Parameters}(b)).

The hexagonal unit cell volume $V$, calculated from our measurements of the lattice parameters, is plotted in Fig.~\ref{Lattice_Parameters}(c) for $T < T_C$, and half the orthorhombic unit cell volume is plotted for $T > T_C$.
This choice was made because the volume of the orthorhombic unit cell is, by definition, exactly twice that of the hexagonal unit cell~\cite{Wilson64,Lazewski11}.
The plot of volume in Fig.~\ref{Lattice_Parameters}(c), therefore, displays the intrinsic discontinuity at $T_C$ in the volumetric thermal expansion.
Upon warming through $T_C$, the volume of our single crystals decreases by $\sim$1.5\%, which is less than the previously reported value~\cite{Willis54} of $\sim$1.9\%.
This discrepancy is related to the fact that, while there is a larger decrease in the basal plane dimensions at $T_C$, it is counterbalanced by an increase in the dimension perpendicular to the basal plane.

The Clausius-Clapeyron relation, $dT_C/dP = \Delta v/\Delta s$, can be used to estimate the initial pressure dependence of $T_C$ using the discontinuities in specific volume, $v$, and specific entropy, $s$, at $T_C$.
$\Delta s$ = 1.77 J mol$^{-1}$ K$^{-1}$ was estimated by numerically integrating $C/T$ data with respect to $T$ (see the inset of Fig.~\ref{Heat_Capacity}) and determining the entropic discontinuity at $T_C$.
The volumetric discontinuity was determined using our XRD data to be $\Delta v = -3.09 \times 10^{-7}$ m$^3$ mol$^{-1}$.
Using these results, the Clausius-Clapeyron relation predicts that $dT_C/dP = -1.75 \times 10^{-7}$ K Pa$^{-1}$ for our single crystals.
This result is within about 12\% of the measured result of $dT_C/dP = -1.56 \times 10^{-7}$ K Pa$^{-1}$~\cite{Goodenough67}, indicating that \textit{both} the ambient-pressure and pressure-dependent properties of the ferromagnetic phase transition in our single crystals agree with results reported in the literature for stoichiometric MnAs samples.

Our XRD measurements were performed to a maximum temperature of 315 K, so the behavior of our single crystals in the vicinity of $T_S$ was not studied.
Previous reports of such measurements show subtle inflections in the temperature dependence of the lattice parameters at $T_S$ ~\cite{Willis54}.
In Fig.~\ref{Lattice_Parameters}(d), the hexagonal (212) Bragg reflection is shown at a temperature (corresponds to right-hand axis) near $T_C$.
For $T > T_C$, this peak is absent and the evolution of the orthorhombic (231), (215), and (421) Bragg reflections with temperature is shown as a function of $2\theta$.
At $T_S$ (not measured), these Bragg reflections are expected to vanish and the (212) hexagonal peak will re-emerge.

\subsection{Mechanism for high-temperature structural transition}


It has been proposed that the mechanism responsible for driving the phase transition at $T_S$ is related to the relaxation of a soft acoustic phonon mode near the $M$ point in reciprocal space~\cite{Lazewski10, Lazewski11}.
Phonon dispersion curves were calculated with spin-polarized density functional theory, observing that a soft phonon mode was favorable for small unit-cell volumes~\cite{Lazewski10, Lazewski11}.
The phonon mode becomes soft as magnetoelastic coupling induces a significant decrease in the unit-cell volume upon warming through $T_C$, and the resulting ionic displacements lower the crystal symmetry from hexagonal to orthorhombic.
As temperature increases further, thermal expansion causes the volume of the unit cell to increase until $T_S$, where the energy associated with the soft phonon mode becomes unfavorably large, the ionic displacements relax, and the hexagonal crystal structure is recovered~\cite{Lazewski10, Lazewski11}.

The low $T_S$ value in our As-deficient samples may imply that the phonon dispersion curves are different from those calculated in Refs.~~\cite{Lazewski10} and ~\cite{Lazewski11} or that the volume at which the soft phonon mode hardens is reached at a lower temperature in our single crystals than that reported for stoichiometric samples of MnAs.
Performing inelastic neutron scattering measurements would be required to experimentally determine the phonon dispersion curves of our samples.
On the other hand, if the critical volume where the soft mode hardens is reached at a lower temperature than usual in our single crystals, then we are able to consider a few possible reasons.
These possibilities include: (1) the volume decrease in our single crystals at $T_C$ is smaller than $\sim$1.9\%, (2) the thermal expansion in the temperature region $T_C < T < T_S$ is anomalously large, and/or (3) the unit cell volume of our single crystals is larger than typical for all temperatures.
As we have already discussed, the volume decrease at $T_C$ of $\sim$1.5\% is lower than previously reported results (see Fig.~\ref{Lattice_Parameters}(c))~\cite{Willis54}, and the negative thermal expansion along [100] in the orthorhombic phase (see Fig.~\ref{Lattice_Parameters}(b)) does not even qualitatively agree with previous results for the thermal expansion for $T_C < T < T_S$.
These observations might be consistent with a scenario in which a ``critical volume'' is reached at a lower temperature.
However, rather than being larger, the volumetric thermal expansion of our single crystals is smaller than previously reported results and the unit-cell volume at room temperature is consistent with other reports.
This implies that there may be other factors besides volume and thermal expansion that should be considered.
One such factor, unit cell anisotropy, is considered next.

\begin{figure}
\begin{center}
\includegraphics[width=\columnwidth]{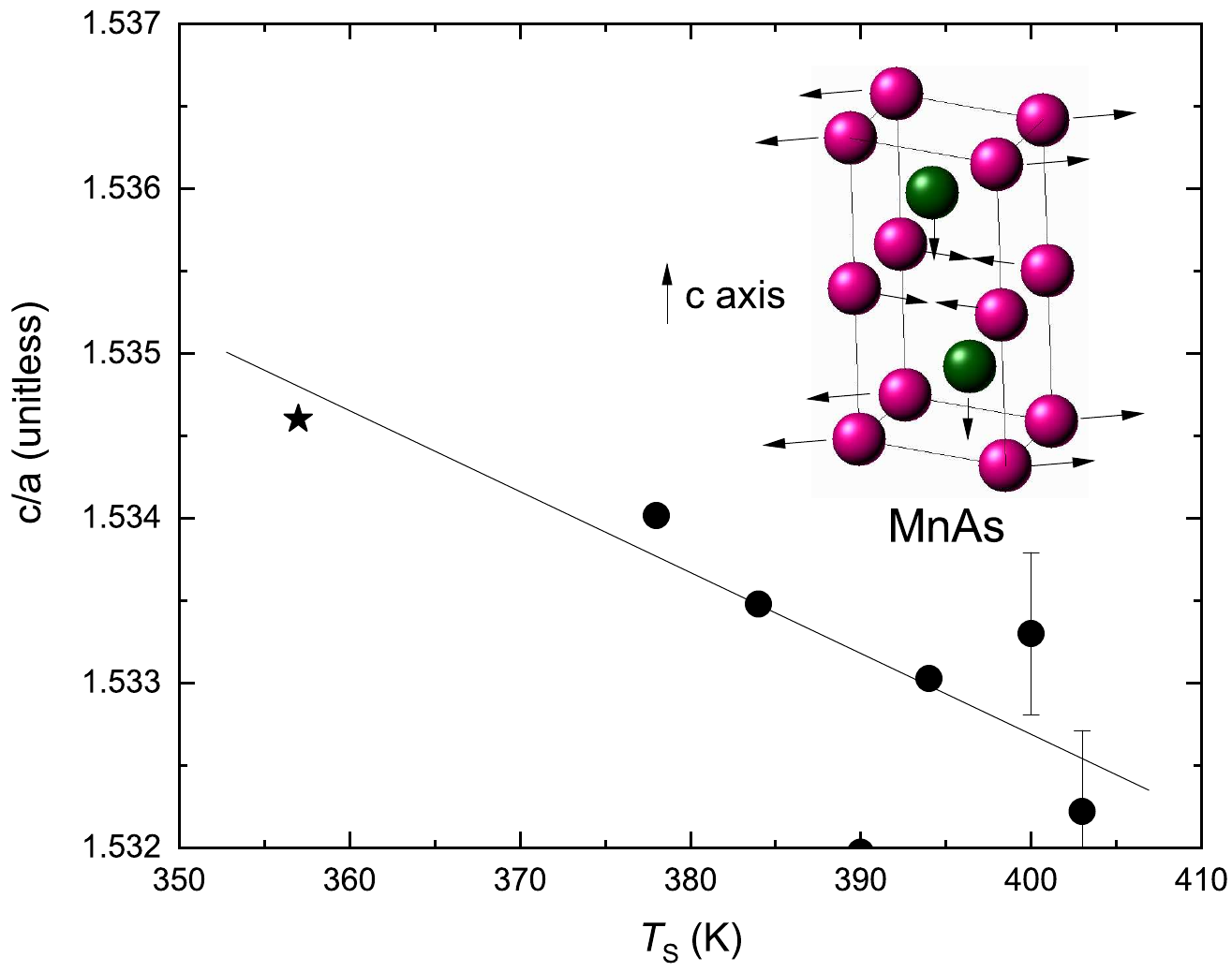}
\caption{\label{Lattice_Anisotropy} (Color online) The ratio of hexagonal lattice parameters, $c/a$, obtained at room temperature for samples in this and other studies, is plotted as a function of the structural phase transition temperature $T_S$ measured using the same samples.
{The result from this study is denoted with a star symbol for clarity.}
Error bars were estimated from uncertainty in lattice parameter values and are smaller than the symbol size in most cases.
The line is a guide to the eye.
The hexagonal unit cell of MnAs is displayed in the inset with the direction of ionic displacements associated with a soft phonon mode explicitly indicated by arrows.
The Mn ions are located on the boundaries of the unit cell and are pink while the As ions reside entirely within the interior of the unit cell and are green.}
\end{center}
\end{figure}

The ionic displacements that lower the symmetry of the crystal structure to orthorhombic for temperatures $T_C \le T < T_S$, and which are associated with a soft phonon mode, are indicated in the inset of Fig.~\ref{Lattice_Anisotropy}.
These displacements involve Mn ions shifting within the hexagonal basal plane and As ions shifting perpendicular to the basal plane~\cite{Lazewski11}.
It is reasonable to expect that these orthogonal ionic displacements might be sensitive to unit-cell anisotropy, which is characterized by the ratio of hexagonal lattice parameters, $c/a$.
To examine this hypothesis, we have plotted values of $c/a$ (obtained at room temperature) as a function of $T_S$ for our single crystals and for other reports from the literature (see Fig.~\ref{Lattice_Anisotropy}).
The data point plotted for $T_S \simeq$ 378 K represents results from a single crystal that we grew using our molten flux technique in a separate batch.
The values of $c/a$ are tabulated in Table~\ref{Table}.
A correlation between $c/a$ and $T_S$ is suggested by Fig.~\ref{Lattice_Anisotropy}.
According to Ref.~~\cite{Goodenough67}, $c/a$ increases with increasing temperature from $c/a$ $\sim$ 1.53 for $T < T_C$ to $c/a$ $\sim$ 1.55 for $T_C < T < T_S$, and finally to $c/a$ $\sim$ 1.56 for $T > T_S$.
We note that for $T_C \le T < T_S$, this anisotropy is characterized by $a_{\mathrm{orth}}/b_{\mathrm{orth}}$.
Given this continuous increase of $c/a$ with temperature (increasing structural anisotropy), perhaps a sample with higher anisotropy at room temperature (higher value of $c/a$) like we have, allows a critical value of $c/a$ (and $T_S$) to be reached at a lower temperature.
In the study by \L a\.{z}ewski \textit{et al.}, a fixed value of $c/a$ = 1.502 was used as the unit cell volume was varied (\textit{i.e.}, simulating the effect of applied pressure) to tune the frequency of the soft phonon mode~\cite{Lazewski11}.
It would be interesting to determine the effect of structural anisotropy on this soft phonon mode.

Based on our results, the volume and its temperature dependence and/or differences in structural anisotropy due to varying As content are responsible for tuning the temperature $T_S$ of the second-order structural phase transition in MnAs.
The value of $T_S$ appears to be decoupled from the magnetic properties since all our results except for $T_S$ agree with results from stoichiometric MnAs samples (see Table~\ref{Table}).
The consequences, if any, of such a low $T_S$ value are unknown.
It would be interesting to determine whether the applied pressure dependence of $T_S$ in our single crystals is different from the established phase diagram in Ref.~~\cite{Menyuk69}.
Given the extreme sensitivity of the transition temperature to structural anisotropy (\textit{i.e.}, a 1\% change in $c/a$ produces a $\sim$10\% change in $T_S$), it would be even more interesting to study the effect of applied uniaxial pressure on $T_S$.

\section{Conclusions}
Measurements of magnetization, specific heat, and thermal expansion were performed on As-deficient MnAs$_{0.968(15)}$ single crystals, which were grown in a molten Sn flux.
Ferromagnetic order with Curie temperatures $T_C \simeq$ 306 K on warming and $T_C \simeq$ 302 K on cooling was observed.
These transition temperatures are consistent with previously-reported values for polycrystalline samples.
In dramatic contrast, the second-order structural phase transition is observed at $T_S \simeq$ 353 K, which is nearly 50 K lower than is typically observed.
In measurements of the thermal expansion of our single crystals, we observed several surprising results that distinguish the behavior of our samples from that reported in other studies on stoichiometric samples.
These results include: (1) a $\sim$1.5\% volume decrease at $T_C$ that is smaller than the reported~\cite{Willis54} value of 1.9\%, (2) a discontinuous jump of $\sim$1.1\% at $T_C$ in the lattice parameters perpendicular to the basal plane instead of being continuous across $T_C$, and (3) negative thermal expansion perpendicular to the basal plane for $T_C \le T \le$ 315 K.

We have considered the observed differences in the behavior of the thermal expansion of our single crystals, relative to previous studies of stoichiometric MnAs, in order to evaluate whether they could be responsible for the anomalously-low value of $T_S$.
This possibility is predicated on a hypothesis that $T_S$ is connected with a critical unit-cell volume at which the soft phonon mode, which is responsible for inducing the orthorhombic crystal structure for $T_C \le T < T_S$, hardens and the hexagonal crystal structure is recovered.
Our thermal expansion results do not appear to be entirely consistent with such a scenario, but we are unable to completely rule out their involvement in tuning $T_S$.
A compelling correlation between the ratio of hexagonal lattice parameters, $c/a$, and $T_S$ suggests that the degree of structural anisotropy in MnAs plays a significant role.
Further study will be needed to determine the extent to which these factors are involved in governing the second-order structural phase transition in MnAs.

\section*{Acknowledgements}
Sample synthesis was funded by the US Air Force Office of Scientific Research - Multidisciplinary University Research Initiative under Grant No. FA9550-09-1-0603.
Measurements of magnetization and specific heat were performed with support from the US Department of Energy, Basic Energy Sciences, under Grant DE-FG02-04ER46105.
Work at Stony Brook University (J. W. S. and M. C. A.) was supported by National Science Foundation Grant No. NSF-DMR-1310008.
The Stony Brook University single crystal diffractometer was obtained through the support of the National Science Foundation Grant CHE-0840483.
Use of the National Synchrotron Light Source, Brookhaven National Laboratory, was supported by the U.S. Department of Energy, Office of Science, Office of Basic Energy Sciences, under Contract No. DE-AC02-98CH10886.
Work at UBC (M. C. A.) was supported by the Natural Sciences and Engineering Research Council of Canada (NSERC).

\section*{References}

\providecommand{\newblock}{}


\begin{thebibliography}{10}
\expandafter\ifx\csname url\endcsname\relax
  \def\url#1{{\tt #1}}\fi
\expandafter\ifx\csname urlprefix\endcsname\relax\def\urlprefix{URL }\fi
\providecommand{\eprint}[2][]{\url{#2}}

\bibitem{Heusler04}
Heusler F 1904 {\em Z. Agnew. Chem.\/} {\bf 17} 260

\bibitem{Bean62}
Bean C~P and Rodbell D~S 1962 {\em Phys. Rev.\/} {\bf 126} 104

\bibitem{Wilson64}
Wilson R~H and Kasper J~S 1964 {\em Acta Cryst.\/} {\bf 17} 95

\bibitem{Campos06}
de~Campos A, Rocco D~L, Carvalho A~M~G, Caron L, Coelho A~A, Gama S, da~Silva
  L~M, Gandra F~C~G, dos Santos A~O, Cardoso L~P, von Ranke P~J and de~Oliveira
  N~A 2006 {\em Nature Mater.\/} {\bf 5} 802

\bibitem{Gama04}
Gama S, Coelho A~A, de~Campos A, Carvalho A~M~G, Gandra F~C~G, von Ranke P~J
  and de~Oliveira N~A 2004 {\em Phys. Rev. Lett.\/} {\bf 93} 237202

\bibitem{Ryu08}
Ryu K~S, Kim J, Lee Y, Akinaga H, Manago T, Viswan R and Shin S~C 2008 {\em
  Appl. Phys. Lett.\/} {\bf 92} 082503

\bibitem{Helman2021}
Helman C, Camjayi A, Islam E, Akabori M, Thevenard L, Gourdon C and Tortarolo M
  2021 {\em Phys. Rev. B\/} {\bf 103} 134408

\bibitem{Wang2019}
Wang B, Zhang Y, Ma L, Wu Q, Guo Y, Zhang X and Wang J 2019 {\em Nanoscale\/}
  {\bf 11} 4204--4209

\bibitem{Saparov12}
Saparov B, Mitchell J~E and Sefat A~S 2012 {\em Supercond. Sci. Technol.\/}
  {\bf 25} 084016

\bibitem{Cheng15}
Cheng J~G, Matsubayashi K, Wu W, Sun J~P, Lin F~K, Luo J~L and Uwatoko Y 2015
  {\em Phys. Rev. Lett.\/} {\bf 114} 117001

\bibitem{Wu14}
Wu W, Cheng J, Matsubayashi K, Kong P, Lin F, Jin C, Wang N, Uwatoko Y and Luo
  J 2014 {\em Nat. Comm.\/} {\bf 5} 5508

\bibitem{Lazewski10}
{\L}a{\.{z}}ewski J, Piekarz P, Tobo{\l}a J, Wiendlocha B, Jochym P~T, Sternik
  M and Parlinski K 2010 {\em Phys. Rev. Lett.\/} {\bf 104} 147205

\bibitem{Lazewski11}
{\L}a{\.{z}}ewski J, Piekarz P and Parlinski K 2011 {\em Phys. Rev. B\/} {\bf
  83} 054108

\bibitem{Barner77}
B{\"{a}}rner K, Braunstein R and Chock E 1977 {\em Phys. Status Solidi B\/}
  {\bf 80} 451

\bibitem{Haneda77}
Haneda S, Kazama N, Yamaguchi Y and Watanabe H 1977 {\em J. Phys. Soc. Jpn.\/}
  {\bf 42} 1201

\bibitem{Campos11}
de~Campos A, Mota M~A, Gama S, Coelho A~A, White B~D, da~Luz M~S and Neumeier
  J~J 2011 {\em J. Cryst. Growth\/} {\bf 333} 54

\bibitem{Blois63JAP}
de~Blois R~W and Rodbell D~S 1963 {\em J. Appl. Phys.\/} {\bf 34} 1101

\bibitem{Blois63PR}
de~Blois R~W and Rodbell D~S 1963 {\em Phys. Rev.\/} {\bf 130} 1347

\bibitem{Paitz71}
Paitz J 1971 {\em J. Cryst. Growth\/} {\bf 11} 218

\bibitem{Zhigadlo17}
Zhigadlo N~D 2017 {\em J. Cryst. Growth\/} {\bf 480} 148--153

\bibitem{Willis54}
Willis B~T~M and Rooksby H~P 1954 {\em Proc. R. Soc. London Sect. B\/} {\bf 67}
  290

\bibitem{Petricek14}
Pet{\v{r}}{\'{i}}{\v{c}}ek V, Du{\v{s}}ek M and Palatinus L 2014 {\em Z.
  Kristallogr.\/} {\bf 229} 345--352

\bibitem{Palatinus07}
Palatinus L and Chapuis G 2007 {\em J. Appl. Cryst.\/} {\bf 40} 786--790

\bibitem{Honda26}
Honda K and Kaya S 1926 {\em Sci. Rep. Tohoku Univ.\/} {\bf 15} 721

\bibitem{Kaya28}
Kaya S 1928 {\em Sci. Rep. Tohoku Univ.\/} {\bf 17} 639

\bibitem{Paige84}
Paige D~M, Szpunar B and Tanner B~K 1984 {\em J. Mag. Mag. Mater.\/} {\bf 44}
  239

\bibitem{Zieba78}
Zieba A, Selte K, Kjekshus A and Andresen A~F 1978 {\em Acta Chem. Scand.\/}
  {\bf 32A} 173

\bibitem{Krokoszinski82}
Krokoszinski H~J, Santandrea C, Gmelin E and B{\"{a}}~rner K 1982 {\em Phys.
  Stat. Sol. (b)\/} {\bf 113} 185

\bibitem{Goodenough67}
Goodenough J~B and Kafalas J~A 1967 {\em Phys. Rev.\/} {\bf 157} 389

\bibitem{Menyuk69}
Menyuk N, Kafalas J~A, Dwight K and Goodenough J~B 1969 {\em Phys. Rev.\/} {\bf
  177} 942

\end{thebibliography}
\end{document}